\documentclass[11pt]{article}
\usepackage{amsmath,amsthm,latexsym,amssymb,amsfonts,epsfig}

\makeatletter
\@addtoreset{equation}{section}
\makeatother

\def\be{\begin{equation}}
\def\ee{\end{equation}}
\def\ba{\begin{eqnarray}}
\def\ea{\end{eqnarray}}

\newcommand\nn{\nonumber}
\newcommand\q{\quad}
\def\Nl{{\mathchoice
{\setbox0=\hbox{$\displaystyle\rm N$}\hbox{\hbox to0pt
{\kern0.4\wd0\vrule height0.9\ht0\hss}\box0}}
{\setbox0=\hbox{$\textstyle\rm N$}\hbox{\hbox to0pt
{\kern0.4\wd0\vrule height0.9\ht0\hss}\box0}}
{\setbox0=\hbox{$\scriptstyle\rm N$}\hbox{\hbox to0pt
{\kern0.4\wd0\vrule height0.9\ht0\hss}\box0}}
{\setbox0=\hbox{$\scriptscriptstyle\rm N$}\hbox{\hbox to0pt
{\kern0.4\wd0\vrule height0.9\ht0\hss}\box0}}}}
\def\Zl{{\mathchoice
{\setbox0=\hbox{$\displaystyle\rm Z$}\hbox{\hbox to0pt
{\kern0.4\wd0\vrule height0.9\ht0\hss}\box0}}
{\setbox0=\hbox{$\textstyle\rm Z$}\hbox{\hbox to0pt
{\kern0.4\wd0\vrule height0.9\ht0\hss}\box0}}
{\setbox0=\hbox{$\scriptstyle\rm Z$}\hbox{\hbox to0pt
{\kern0.4\wd0\vrule height0.9\ht0\hss}\box0}}
{\setbox0=\hbox{$\scriptscriptstyle\rm Z$}\hbox{\hbox to0pt
{\kern0.4\wd0\vrule height0.9\ht0\hss}\box0}}}}
\def\Ql{{\mathchoice
{\setbox0=\hbox{$\displaystyle\rm Q$}\hbox{\hbox to0pt
{\kern0.4\wd0\vrule height0.9\ht0\hss}\box0}}
{\setbox0=\hbox{$\textstyle\rm Q$}\hbox{\hbox to0pt
{\kern0.4\wd0\vrule height0.9\ht0\hss}\box0}}
{\setbox0=\hbox{$\scriptstyle\rm Q$}\hbox{\hbox to0pt
{\kern0.4\wd0\vrule height0.9\ht0\hss}\box0}}
{\setbox0=\hbox{$\scriptscriptstyle\rm Q$}\hbox{\hbox to0pt
{\kern0.4\wd0\vrule height0.9\ht0\hss}\box0}}}}
\def\Rl{{\mathchoice
{\setbox0=\hbox{$\displaystyle\rm R$}\hbox{\hbox to0pt
{\kern0.4\wd0\vrule height0.9\ht0\hss}\box0}}
{\setbox0=\hbox{$\textstyle\rm R$}\hbox{\hbox to0pt
{\kern0.4\wd0\vrule height0.9\ht0\hss}\box0}}
{\setbox0=\hbox{$\scriptstyle\rm R$}\hbox{\hbox to0pt
{\kern0.4\wd0\vrule height0.9\ht0\hss}\box0}}
{\setbox0=\hbox{$\scriptscriptstyle\rm R$}\hbox{\hbox to0pt
{\kern0.4\wd0\vrule height0.9\ht0\hss}\box0}}}}
\def\Cl{{\mathchoice
{\setbox0=\hbox{$\displaystyle\rm C$}\hbox{\hbox to0pt
{\kern0.4\wd0\vrule height0.9\ht0\hss}\box0}}
{\setbox0=\hbox{$\textstyle\rm C$}\hbox{\hbox to0pt
{\kern0.4\wd0\vrule height0.9\ht0\hss}\box0}}
{\setbox0=\hbox{$\scriptstyle\rm C$}\hbox{\hbox to0pt
{\kern0.4\wd0\vrule height0.9\ht0\hss}\box0}}
{\setbox0=\hbox{$\scriptscriptstyle\rm C$}\hbox{\hbox to0pt
{\kern0.4\wd0\vrule height0.9\ht0\hss}\box0}}}}
\def\Hl{{\mathchoice
{\setbox0=\hbox{$\displaystyle\rm H$}\hbox{\hbox to0pt
{\kern0.4\wd0\vrule height0.9\ht0\hss}\box0}}
{\setbox0=\hbox{$\textstyle\rm H$}\hbox{\hbox to0pt
{\kern0.4\wd0\vrule height0.9\ht0\hss}\box0}}
{\setbox0=\hbox{$\scriptstyle\rm H$}\hbox{\hbox to0pt
{\kern0.4\wd0\vrule height0.9\ht0\hss}\box0}}
{\setbox0=\hbox{$\scriptscriptstyle\rm H$}\hbox{\hbox to0pt
{\kern0.4\wd0\vrule height0.9\ht0\hss}\box0}}}}
\def\Ol{{\mathchoice
{\setbox0=\hbox{$\displaystyle\rm O$}\hbox{\hbox to0pt
{\kern0.4\wd0\vrule height0.9\ht0\hss}\box0}}
{\setbox0=\hbox{$\textstyle\rm O$}\hbox{\hbox to0pt
{\kern0.4\wd0\vrule height0.9\ht0\hss}\box0}}
{\setbox0=\hbox{$\scriptstyle\rm O$}\hbox{\hbox to0pt
{\kern0.4\wd0\vrule height0.9\ht0\hss}\box0}}
{\setbox0=\hbox{$\scriptscriptstyle\rm O$}\hbox{\hbox to0pt
{\kern0.4\wd0\vrule height0.9\ht0\hss}\box0}}}}
\title{Are the spectra of geometrical operators in Loop Quantum Gravity
really discrete?}

\author{Bianca Dittrich$^1$\thanks{bdittrich AT perimeterinstitute.ca  }  , 
Thomas Thiemann$^{1,2}$\thanks{Thomas.Thiemann AT aei.mpg.de   }\\
$^1$ Perimeter Institute for Theoretical Physics \\
31 Caroline Street North, Waterloo, ON N2L 2Y5, Canada\\
$^2$ MPI f. Gravitationsphysik, Albert-Einstein-Institut, \\
    Am M\"uhlenberg 1, 14476 Potsdam, Germany
}

\begin{document}

\maketitle

\begin{abstract}
One of the celebrated results of Loop Quantum Gravity (LQG) is the 
discreteness of the spectrum of geometrical operators such as length, 
area and volume operators. This is an indication that Planck scale 
geometry in LQG is discontinuous rather than smooth. However, there is 
no rigorous proof thereof at present, because the afore mentioned 
operators are not gauge invariant, they do not commute with the quantum 
constraints. 

The relational formalism in the incarnation of Rovelli's partial and 
complete observables provides a possible mechanism for turning a non 
gauge invariant operator into a gauge invariant one.

In this paper we investigate whether the spectrum of such a physical, that is 
gauge invariant, observable can be predicted from the spectrum of the 
corresponding gauge variant observables. We will not do this in full 
LQG but rather consider much simpler examples where field theoretical 
complications are absent. We find, even in those simpler cases, that 
kinematical discreteness of the spectrum does not necessarily survive 
at the gauge invariant level. Whether or not this happens depends 
crucially on how the gauge invariant completion is performed. 

This indicates that ``fundamental discreteness at Planck scale in LQG'' 
is far from established.   
To prove it, one must provide the detailed 
construction of gauge invariant versions of geometrical operators.   
\end{abstract}

\section{Introduction}

 Among the candidates for a theory of quantum gravity Loop Quantum 
 Gravity (LQG) \cite{lqgreviews} has gained more and more popularity in 
 recent years. One of the successes of LQG is the rigorous construction 
 of (spatial) geometrical operators, such as the area, the volume and 
the length 
 operator \cite{geomop,geomop2} on the so called kinematical Hilbert space of 
 LQG \cite{AIL}. Moreover it turns out that these geometrical operators 
 have a discrete spectrum. In particular there exists an area gap 
 \cite{geomop,geomop2} but apparently no volume gap 
 \cite{johannes}.\footnote{There exist a volume gap on the subspace of the kinematical Hilbert space spanned by spin networks with vertex valence less or equal to four. However, on the full kinematical Hilbert space numerical evidence for an accumulation point at zero was found, see  the last three references in \cite{johannes}.} 

 Another result is \cite{lost}, which shows that the kinematical Hilbert 
 space of LQG is the unique quantum representation for the holonomies 
 and flux variables used in LQG satisfying certain covariance conditions 
 with respect to the (spatial) diffeomorphism group. This shows that the 
 discreteness of the spectra for the geometrical operators follows from 
 a minimal set of requirements. Furthermore the volume operator is 
pivotal to obtain a uv--finite 
 quantization of the (gravity and matter) Hamiltonian constraints of 
the theory \cite{hamilt}. 

 In summary the geometrical operators play a  
 very important role for the structure and further developement of 
 LQG. However, so far the discreteness of their spectrum is a result at 
the kinematical level only. With 
 'kinematical level' we mean that the geometrical operators defined so 
 far are gauge dependent, i.e. they are not invariant under 
 spacetime diffeomorphisms. True observables have to be gauge 
independent, in the 
 canonical formalism such observables are known as Dirac observables. 
 Physical measurements are described by Dirac observables and not by 
 kinematical observables. 

 The explicit construction of Dirac observables in general relativity is 
 very difficult since it requires the solution of the dynamics of the 
 theory. However methods to obtain Dirac observables (at least as 
 classical phase space functions) are available \cite{bd, reduced}. With 
 these methods one can construct so--called complete observables 
 \cite{Rov}, a specific type of relational observables. The main idea 
 here is to use some set of fields $T$ as ``clocks and rods'' and to 
 express another field $f$ (the ``partial observable'') in the 
 coordinates defined by these ``clocks and rods''. Under a 
 diffeomorphism both $f$ and $T$ transform in the same way, so that the 
 relation between $f$ and $T$ does not change. More details and a proof 
 that the resulting phase space functions are indeed Dirac observables 
 can be found in the second reference in \cite{bd}.

 Indeed one way how to construct Dirac observables out of the 
 geometrical operators mentioned above is to use matter fields in order 
 to localize the region of which one wants to measure for instance the 
 (spatial or space--time) volume. This is also suggested in \cite{geomop}. The question would be if the Dirac 
 observables constructed in this manner (assuming that such Dirac 
 observables can be quantized as self--adjoint operators on a yet to be 
 constructed physical Hilbert space) have a discrete spectrum or not. \cite{geomop} argues for a discrete spectrum of the Dirac observables.

 Contrary to this expectation we will show that in this note, without further assumptions, it is 
impossible to 
 make any predictions about the spectra of the complete observables, 
 even if the spectra of the partial observables and the clocks are 
 known. We will do that by means of quite simple examples with finitely 
many degrees of freedom, where the 
 physical Hilbert space can be explicitly constructed. In more 
 complicated (field theory) examples, even the physical Hilbert space 
might not be 
 unique and the spectra of observables could in principle depend on the 
 choice of the Hilbert space.\\
\\
The plan of the paper is as follows:
 
 In the next section we will explain the method of partial and complete 
 observables. This allows us to associate to a gauge variant function 
 (the partial observable) a family of gauge invariant functions (the 
 complete observable). Futhermore one has to choose a clock. We can then 
 consider in which way the spectrum of the complete observables depends 
 on the spectrum of the partial observable, the clock and the 
constraint. 
 
 In section \ref{examples} we will show with the help of concrete 
 examples that {\it every possible combination of continuous and 
discrete 
 spectra for the constraint, the partial observable, the clock and the 
 complete observable can be realized}. 

We will end with a discussion of the results and their meaning for LQG.
In particular we will coment on earlier results within the LQG literature 
on the physical spectrum of geometrical operators in the context of 2+1 
Lorentzian gravity \cite{etera,noui} and 3+1 Lorentz covariant 
gravity \cite{alexandrov}.

In an appendix we consider an additional 
toy model which is geared to constructing a gauge invariant 
observable from a baby version of a volume operator by means of a baby 
version of a scalar field as discussed in \cite{geomop}. We find that 
in contrast to the arguments spelled out there, the kinematically 
discrete spectrum switches to a gauge invariant continuous one.

\section{Partial and Complete Observables} \label{section2}

Here we will explain the method of partial and complete observables. The examples we are going to discuss exhibit only one constraint, i.e. only one gauge degree of freedom. We will comment on the case with several constraints (general relativity has infinitely many constraints) lateron.

Let us start with an example, namely the relativistic particle in two space--time dimensions. Here we have the mass shell constraint
\ba\label{massshell}
C=p_0^2-p_1^2-m^2
\ea 
where $m$ is the mass and $(p_0,p_1)$ the momentum of the particle. Now we have to choose a clock variable $T$ and another partial observable $f$. Our choice is $T=x_0$ and $f=x_0 p_1-x_1p_0$ where $(x_0,x_1)$ is the position vector of the particle. The complete observable $F(\tau)$, where $\tau$ is a phase space independent parameter, is defined to be the gauge invariant phase space function with the following property: On the hypersurface $T=\tau$ it should coincide with the phase space function $f$. Gauge invariance means here that $F(\tau)$ has to be invariant under the flow of the mass shell constraint (\ref{massshell}), i.e. that $\{F(\tau),C\}=0$, where the Poisson brackets are defined by $\{x_i,p_j\}=\delta_{ij}$.

One way to find the complete observable is to compute the flow of the clock variable $T$ under the constraint $C$, that is
\ba
T(t):=\sum_{k=0}^\infty \frac{t^k}{k!}\{T,C\}_k    
\ea
where $\{g,C\}_{k+1}=\{\{g,C\}_{k},C\},\,\,\{g,C\}_0=g$ are iterated Poisson brackets. Next one has to solve the equation $T(t)\stackrel{!}{=}\tau$ for $t$. The solution $t_s=t_s(\tau)$ is in general also phase space dependent. For the relativistic particle we have
\ba
T(t)=x_0+2t p_0\stackrel{!}{=}\tau \,\, ,
\ea
hence $t_s=\tfrac{1}{2}p_0^{-1}(\tau-x_0)$.

Now we have to insert the solution $t_s$ into the flow of $f$ under the constraint $C$. This gives the complete observable $F(\tau)$:
\ba
F(\tau)=f(t_s(\tau))=\sum_{k=0}^\infty \frac{(t_s(\tau))^k}{k!}\{f,C\}_k   \q .
\ea
Since for the relativistic particle $f(t)=(x_0+2t p_0)p_1 -(x_1-2tp_1)p_0$ we find
\ba\label{boost}
F(\tau)=\tau p_1-(x_1p_0-(\tau-x_0)p_1)=2 \tau p_1 -(x_0p_1+x_1p_0) \q .
\ea
Indeed (\ref{boost}) coincides with $f$ on the hypersurface $T=x^0=\tau$ and is invariant under the flow of $C$.

This construction can be generalized to systems with arbitrary many (first class) constraints by introducing as many clock variables $T_i$ as there are constraints $C_i$, where $i$ is in some index set $I$. The complete observable $F(\tau_i)$ associated to these clock variables and a partial observable $f$ is defined to be the gauge invariant function which coincides on the hypersurface $\{T_i=\tau_i \;\forall i\in I\}$ with the function $f$. For methods to compute complete observables see \cite{bd}. 

We want to remark that the complete observable might not always be well defined. For instance the surface $T=\tau$ might meet a gauge orbit several times which leads to multiple solutions of the equation $T(t)\stackrel{!}{=}\tau$ and hence might lead to multiple valued complete observables as is discussed in the first reference in \cite{bd}. Some examples in the next section will have multiple solutions $t_s$ of the equation $T(t)\stackrel{!}{=}\tau$ but in these the partial observable $f$ is chosen such that $f(t)$ coincides on all solutions $t_s$ and hence we will always get a well defined complete observable. 

On the other hand it might happen that for a fixed parameter $\tau$ the surface $T=\tau$ does not meet a certain set of gauge orbits at all. Here we will assume that the parameter $\tau$ is always chosen such that 
one can define the complete observable at least on an open set of the phase space. In our examples this complete observable can always be continued analytically to a gauge invariant function defined on the whole phase space. 
\footnote{This convention is neccessary for instance for the examples with the rotation generator as constraint. Note that with this convention it may happen that the complete observable $F(\tau)$ assumes negative values whereas the partial observable $f$ is non--negative.}

Furthermore often we will give the complete observables modulo terms that vanish on the constraint hypersurface, since these terms are not relevant for the quantization of the observables on the physical Hilbert space.

With the relativistic particle we already have an example where the spectrum of the quatization of the partial observable $f=x_0p_1-x_1p_0$ is discrete (since it generates rotations in the $(x_0,x_1)$--plane) and the spectrum of the quantization of the corresponding complete observable $F(\tau=0)=-x_0p_1-x_1p_0$ is continuous (since it generates boosts).

Such a change of the spectral properties can be understood in the following way: 

Heuristically the quantisation $\hat f$ of a (smooth) phase space function $f$ has a discrete spectrum or a continuous spectrum if the vector field $\chi_f$ associated to the phase space function $f$ generates compact orbits or non--compact orbits respectively.    

The vector field $\chi_f$ associated to $f$ is defined by the equation
\ba
\{g,f\}=\chi_f(g)
\ea
which has to hold for every smooth function $g$. Here the Poisson brackets $\{\cdot,\cdot\}$ are defined via the symplectic form $\omega$ defined on the phase space in question. 

Assume that the equation $T=\tau$ defines a good gauge fixing, i.e. that the surface $T=\tau$ meets each gauge orbit once and only once. As it is explained in \cite{bd,reduced} in this case the following holds: The space of all complete observables with this choice of clock variable $T$ can be mapped via a symplectomorphism onto the reduced phase space obtained with the gauge fixing $T=\tau$ and equipped with the Dirac bracket $\{\cdot,\cdot\}^D$. This map is given by mapping the complete observable associated to $f$ to the gauge restriction of $f$, i.e. the restriction of the function $f$ to the reduced phase space $\{C=0,T=\tau\}$. 

Hence considering complete observables and the symplectic flows they generate is equivalent to considering the reduced phase space $\{C=0,T=\tau\}$ with its induced symplectic form, which is given by the Dirac bracket. That is a complete observables associated to a function $f$ generates compact orbits if and only if this is the case for the symplectic flow of the gauge restriction of $f$ defined via the Dirac bracket. This symplectic flow preserves in particular the reduced phase space $\{C=0,T=\tau\}$. 
In fact the vector fields $\chi^D_f$ associated to the gauge restriction of $f$ via the Dirac bracket can be understood as the vector fields $\chi_f$ projected to the hypersurface $\{C=0,T=\tau\}$. 

Now whether the orbits generated by the vector field $\chi_f$ are compact or not is a global property of the vector field and it might change under the projection onto the reduced phase space. Also the global properties of the reduced phase space and how the reduced phase space is embedded into the full phase space is important. This explains the different spectra for the partial observable $f$ and the complete observable $F(\tau)$. Moreover this argument shows that it is very difficult to predict which kind of spectrum the complete observables will have, since it requires control of the global features of the reduced phase space.

In the next section we will give examples for all possible combinations of discrete (i.e. pure point) or continuous (i.e. absolutely continous) spectra for the (quantized) constraint, the clock variable $T$, the partial observable $f$ and the complete observable $F(\tau)$.

\section{The examples}\label{examples}

In our examples we will use the phase space $\Rl^2 \times \Rl^2$ with canonical coordinates $(x_1,p_1,x_2,p_2)$ and Poisson brackets $\{x_i,p_j\}=\delta_{ij}$ (where $\delta_{ij}$ is the Kronecker symbol). 

We choose the space of square integrable functions $L_2(\Rl^2)$ as the kinematical Hilbert space. The quantized configuration coordinate functions $\hat{x}_j$ act as multiplication operators $\hat{x}_j=x_j$ and the quantized momenta as derivative operators $\hat{p}_j=-i\hbar \partial_j$.

We will start our considerations with the constraint $\hat C_d= \hat x_1 \hat p_2-\hat x_2 \hat p_1$, that is the generator of rotations on $\Rl^2$. The constraint operator has a discrete spectrum $\{ \hbar z, z\in \Zl\}$ (indicated by the index $d$). We will assume that the partial and complete observables have either an entirely discrete (i.e. pure point) or entirely continuous (i.e. absolutely continuous) spectrum and that the spectral type of the complete observable does not depend on the particular value of the parameter $\tau$. Then there exist eight combinations for the spectra of the clock variable $\hat T$ the partial observable $\hat f$ and the complete observable $\hat{F}(\tau)$. We will indicate these combinations by 
(d,i,j,k) with $\text{i,j,k}=\{\text{c,d}\}$ for the spectrum i of the clock variable $\hat T$, the spectrum j of the partial observable $\hat{f}$ and the spectrum k of the complete observable $\hat{F}(\tau)$.




The construction of the physical Hilbert space for the constraint $\hat C_d$ is straightforward.

The constraint has a pure point spectrum and moreover the point zero is included in this spectrum so that we have solutions $\psi_{phys}$ to the constraint equation $\hat C_d \psi_{phys}=0$ that are elements of the kinematical Hilbert space. A physical inner product is given by the restriction of the kinematical inner product to these solutions. 

The gauge invariant functions we will encounter in our examples of complete observables are given by $r^2:=x_1^2+x_2^2$ and $h:=x_1^2+x_2^2+p_1^2+p_2^2$. The corresponding quantizations have continuous spectrum and discrete spectrum respectively. Details on the physical Hilbert space and the algebra of Dirac observables can be found in the appendix \ref{appendix}.

We will begin the discussion of the examples with the case (d,d,d,d). Here we choose as clock variable and as partial observable
\ba
T= x_1^2+p_1^2 \q ,     \q\q\q f=x_2^2+p_2^2   \q ,
\ea
that is the harmonic oscillators in the $x_1$-- and $x_2$--coordinates respectively. The corresponding quantum operators have obviously discrete spectrum.

To find the complete observable we have first to solve the equation
\ba
T(t):=\sum_{k=0}^\infty \frac{t^k}{k!}\{T, C_d\}_k\stackrel{!}{=}\tau
\ea
for $t$ and then to insert the solutions of this equation into $f(t)$. However there is a shorter way: Note that
\ba
T(t) &=&  x_1^2 \cos^2(t)+x_2^2 \sin^2(t)-2x_1x_2\cos(t)\sin(t)+ \nn\\
&&p_1^2\cos^2(t)+p_2^2\sin^2(t)-2p_1p_2\cos(t)\sin(t)  \q .
\ea
Hence we have for solutions $t_s$ of the equation $T(t)=\tau$
\ba
2x_1x_2 \cos(t_s)\sin(t_s)+2p_1p_2\cos(t_s)\sin(t_s)
&=&
x_1^2\cos^2(t_s)+x_2^2\sin^2(t_s)+ \nn\\
&& p_1^2\cos^2(t_s)+p_2\sin^2(t_s)\;-\tau  \q .
\ea
We use this equation in 
\ba
F(\tau)=f(t_s)
&=& 2x_1x_2 \cos(t_s)\sin(t_s)+2p_1p_2\cos(t_s)\sin(t_s)+ \nn\\
&& x_2^2\cos^2(t_s)+x_1^2\sin^2(t_s)+p_2^2\cos^2(t_s)+p_1\sin^2(t_s)
\ea
and find
\ba
F(\tau)=x_1^2+x_2^2+p_1^2+p_2^2-\tau
\ea
that is the sum of two harmonic oscillator Hamiltonians minus a constant $\tau$. The corresponding quantum operator has a discrete spectrum.

Hence we found an example for the case (d,d,d,d).
Examples for all the other cases (d,i,j,k) can be found in a similar manner. The examples are summarized in table (\ref{table1}).

\ba \label{table1}
\begin{array}{|c|c|c|c|}
\hline \text{Case} & T &  f & F(\tau)\\
\hline
\hline 
         \text{ (d,d,d,d)}
        & x_1^2+p_1^2 
        & x_2^2+p_2^2   
        &  x_1^2+p_1^2+x_2^2+p_2^2-\tau \\ 
\hline
\hline  
        \text{(d,c,d,d)}  
        &  (1-\omega^2) x_1^2 + 
        &    p_1^2 + \omega^2 x_1^2   +
        &      x_1^2+p_1^2+x_2^2+p_2^2-\tau      \\
        \underset{}{}
        &  (1-\omega'^2) x_2^2
        &     p_2^2 + \omega'^2 x_2^2
        &         \\ 
\hline
\hline    
         \text{(d,d,c,d)}  
          &   p_1^2 + \omega^2 x_1^2+
        &      (1-\omega^2) x_1^2 + 
       &         x_1^2+p_1^2+x_2^2+p_2^2-\tau  \\
       \underset{}{}
       &     p_2^2 + \omega'^2 x_2^2
        &    (1-\omega'^2) x_2^2
        &           \\
\hline
\hline  
   \text{(d,c,c,d)}
           & x_1^2 +p_2^2  
           & x_2^2+p_1^2 
           & x_1^2+p_1^2+x_2^2+p_2^2-\tau         \\

\hline
\hline   
         \text{(d,d,d,c)}
          & ( p_1^2 + \omega^2 x_1^2)   - 
          & -( p_1^2 + \omega'^2 x_1^2)+ 
         & (\omega^2-\omega'^2) (x_1^2 + x_2^2) -\tau      \\
          \underset{}{}
         & ( p_2^2 + \omega'^2 x_2^2)
         &  ( p_2^2 + \omega^2 x_2^2) 
         &       \\
\hline
\hline    
        \text{(d,d,c,c)}
         & p_1^2+x_1^2 
         & p_1^2-x_2^2
         &  -x_1^2-x_2^2+\tau    \\
          
\hline 
\hline   
           \text{(d,c,d,c)}
          &  p_1^2-x_2^2
          &  p_1^2+x_1^2 
           & x_1^2+x_2^2+\tau      \\
      
\hline
\hline    
          \text{(d,c,c,c)}
         & x_1^2
         & x_2^2 
         & x_1^2+x_2^2-\tau   \\
 
\hline
\hline
\end{array}
\ea   

We will now turn to a constraint with a continuous spectrum. We use the same phase space and the same kinematical Hilbert space as before, namely $\Rl^2\times \Rl^2$ and $L_2(\Rl^2)$ respectively. We will work with the momentum constraint $C_c=p_1$, whose quantization $\hat C_c=-i\hbar\partial_1$ has a continuous spectrum. 

The Dirac observable algebra for this constraint is spanned by $x_2$ and $p_2$, i.e. it is given by phase space functions which do not depend on $x_1$ or $p_1$. Physical wave functions are wave functions which do not depend on $x_1$. A physical inner product can be defined by omitting the integration over $x_2$ in the kinematical inner product. Hence we are left with a physical Hilbert space $L_2(\Rl, dx_2)$ which carries the usual representation of the basic Dirac observables $\hat x_2$ and $\hat p_2$.

The calculation of the complete observables are straightforward. In the following table  (\ref{table2}) we have listed the examples for all possible cases (c,i,j,k) where again i,j,k can take values d or c and indicate a discrete or continuous spectrum  respectively of the corresponding quantum observable.

\ba \label{table2}
\begin{array}{|c|c|c|c|}
\hline \text{Case} & T &  f & F(\tau)\\
\hline
\hline   
\text{(c,d,d,d)}
        & x_1^2+p_1^2 
        & x_1^2+p_1^2+x_2^2+p_2^2 
        &  x_2^2+p_2^2+\tau    \\
\hline\hline 
\text{(c,c,d,d)} 
        &  x_1^2
        &  x_1^2+p_1^2+x_2^2+p_2^2
        &   x_2^2+p_2^2+\tau 
              \\
\hline \hline        
\text{(c,d,c,d)}  
          &   x_1^2+p_1^2+x_2^2+p_2^2 
        &      x_1^2
        &     -x_2^2-p_2^2+\tau              \\

\hline\hline 
\text{(c,c,c,d)}
           & x_1+p_2^2 
           & x_1-x_2^2 
           &  -x_2^2-p_2^2+\tau         \\
\hline \hline  
   \text{(c,d,d,c)}
          &  x_1^2+p_1^2
          &  (x_1p_2-x_2p_1)^2 
         &   \tau p_2^2   \\        
\hline\hline   
\text{(c,d,c,c)}
         & x_1^2+p_1^2+x_2^2+p_2^2 
         & x_1^2+x_2^2
         &  -p_2^2+\tau  \\         
\hline \hline   
\text{(c,c,d,c)}
          &  x_1^2+x_2^2
          &  x_1^2+p_1^2+x_2^2+p_2^2  
           &  p_2^2+\tau  \\
\hline \hline   
\text{(c,c,c,c)}
         & x_1
         & x_1+x_2 
         &  x_2+\tau \\

\hline
\hline 
\end{array}
\ea   

Consider in particular the cases (c,c,d,d) and (c,c,d,c) which can be generalized to 
\ba \label{last1}
T&=&x_1^2+\gamma x_2^2 \nn\\
f&=&x_1^2+p_1^2+x_2^2+p_2^2 \nn\\
F(\tau)&\simeq&(1-\gamma)x_2^2+p_2^2+\tau
\ea
with $0\leq\gamma <1$ or $\gamma \geq 1$ for a discrete spectrum of the complete observable or a continuous spectrum of the complete observable respectively. The symbol $\simeq$ indicates that the last equation in (\ref{last1}) holds modulo terms which vanish on the constraint hypersurface. Hence we have an example where just changing the clock variable $T$ leads to the change of the spectrum of the complete observable.

Examples with more than one constraint can be easily constructed by taking ``a direct sum'' of the above examples for one constraint: For instance consider the constraints $C_1=p_1$ and $C_2=p_3$ on a phase space $\Rl^4\times\Rl^4$. Choose as partial observables
\ba
 T_1&=&x_1^2 \nn\\
 T_2&=&x_3^2  \nn\\
f&=&x_1^2+p_1^2+x_2^2+p_2^2+x_3^2+p_3^2+x_4^2+p_4^2  \q .
\ea
 The corresponding complete observable is $F(\tau_1,\tau_2)\simeq 
 x_2^2+p_2^2+x_4^2+p_4^2+\tau_1+\tau_2$. Hence we have an example with 
 two constraints and two clock variables with continuous spectrum and a 
 partial observable $f$ and a complete observable $F(\tau_1,\tau_2)$ 
 with discrete spectrum.

\section{Discussion}

 The examples show that it is very difficult to make any predictions 
 about the spectra of physical observables even if the spectra of gauge 
 variant observables are known to which the physical observables are 
 associated in a certain sense (e.g. here through the method of partial 
 and complete observables). The examples considered here are very simple 
 and do not include further complications one would expect for the 
 geometrical operators in LQG. For instance one might consider 
 space--time geometrical operators rather than spatial ones.

Among earlier work on these issues we select the following three:
In \cite{etera} the authors consider 2+1 Lorentzian gravity without 
matter and construct a length operator within the Hamiltonian theory. 
Since the authors do not require that the spatial slices be spacelike,
the length operator measures the length of timelike or spacelike curves.
The spacelike spectrum is continuous while the timelike one is discrete 
corresponding to the coexistence of continuous and discrete spectrum of 
the Casimir for SO(1,2). However, the spectrum is at the kinematical level 
only and it is unknown what happens when one implements the dynamics.
Of more relevance to our paper is \cite{noui} which concerns Riemannian 2+1 gravity 
with point particles. Now the length operator can be made into a Dirac observable and 
the spectrum is discrete both for the kinematical and physical length observable! However, it is not clear whether a similar 
construction will also work in 3+1 dimensions, and the analysis is done for the Riemannian case only. Finally there is a body 
of literture called ``Lorentz covariant gravity'' in 3+1 dimensions, see for instance \cite{alexandrov}, 
where one keeps the connection a Lorentz connection at the price of
the connections not to Poisson commute. Hence one does not have any
(connection) representation and thus rigorous spectral theory cannot 
be performed.     
 
 Our heuristic argument 
 of how one could understand the change from discrete to continuous 
 spectra and vice versa shows that the global features of the 
 kinematical and reduced phase space play an important role. However it 
 is very hard to get complete control over the reduced phase 
 space of general relativity. It might be therefore complicated to give 
 any conditions that would ensure the discreteness of the spectra for 
 the complete observables. 

 As a more general lesson these examples show that physical observables 
 might have very unexpected properties. Here we considered the spectra 
 of physical observables. Other properties are for instance commutation 
 relations of physical observables, see for instance the third reference 
 in \cite{bd}. Moreover there are indications that considering physical 
 observables might lead to a notion of non--commutative space--time 
 \cite{fredenhagen}.

 Although we cannot say yet whether the true physical geometrical 
 operators of LQG will have discrete spectra or not without spelling 
out additional details, we want to emphasize that their kinematical 
versions are still important. For instance, the kinematical volume 
operator enters the definition of the Hamiltonian constraint and surely
many other Dirac observables which are gauge invariant aggregates built 
out of gauge variant operators.

\begin{appendix}

\section{Example: Physical Hilbert space} \label{appendix}

 Here we will shortly discuss the representation of the Dirac observable 
 algebra on the physical Hilbert space for the constraint $\hat C_d=\hat 
x_1 \hat p_2-\hat x_2 \hat p_1 $.
The Dirac observable algebra is spanned by
\ba \label{sl2rrep1}
\hat d= \tfrac{1}{2}(\hat x_i\hat p_i+\hat p_i \hat x_i)=\hat x_i  \hat p_i - i \hbar \q\q\q\q
\hat e^+=\hat x_i \hat x_i \quad \q\q\q \hat e^-=\hat p_i \hat p_i   \q  
\ea
where we sum over repeated indices.
With this operator ordering for $\hat d$ the observables (\ref{sl2rrep1}) represent the $sl(2,\Rl)$-algebra:
\ba
[\hat d, \hat e^\pm]=\mp 2i\hbar e^\pm  \q\q\q\q \q\q \q \quad [\hat e^+,\hat e^-]=4i\hbar\hat d   \q  .
\ea


The constraint has a pure point spectrum and moreover the point zero is included in this spectrum so that we have solutions $\psi_{phys}$ to the constraint equation $\hat C_d \psi_{phys}=0$ that are elements of the kinematical Hilbert space. A physical inner product is given by the restriction of the kinematical inner product to these solutions. Explicitly we can change to polar coordinates $r\in \Rl_+, \theta \in [0,2\pi)$:
\ba
x_1=r \cos \theta \q    x_2=r\sin\theta 
\ea
so that the constraint operator $\hat C_d$ and the Dirac observable algebra (\ref{sl2rrep1}) become
\ba\label{physrep}
\hat C_d &=& -i\hbar(x_1\partial_{2}-x_2\partial_{1}) =-i\hbar \partial_\theta \nn\\
\hat d &=&-i\hbar(x_1\partial_1+x_2\partial_2) =-i\hbar(r\partial_r+1) \nn\\
\hat e^+ &=& (x_1^2+x_2^2)=r^2 \nn\\
\hat e^- &=& -\hbar^2(\partial_1^2+\partial^2)=-\hbar^2(\partial_r^2+r^{-1}\partial_r+r^{-2}\partial^2_{\theta})
\ea
on the kinematical Hilbert space $L_2(r d r\,d \theta, \Rl_+\times [0,2\pi))$ .The physical states are the states which do not depend on the angular coordinate $\theta$ and the physical inner product can be written as
\ba
(\psi_{phys},\phi_{phys})_{phys}=\int_{\Rl_+} \overline{\psi_{phys}(r)} \phi_{phys}(r) \, r dr \q .
\ea

The representation (\ref{physrep}) of the Dirac observable algebra (\ref{sl2rrep1}) is unitary and irreducible. It is equivalent to the representation $D^+_{1/2}$  (see for instance \cite{basu}) fom the discrete series of representations of $sl(2,\Rl)$. In this representation the operators $\hat e^+$ and $\hat e^+ + \hat e^-$  (which is the two--dimensional
 harmonic oscillator restricted to states with vanishing angular momentum) have absolutely continuous
 and pure point spectrum respectively. Theses Dirac observables appear as complete observables in our examples.

\section{Baby Version of LQG Geometrical Operators}

The examples considered so far could be taken as irrelevant for the 
situation in LQG for two reasons: First, the configuration space of the 
models was non 
compact while for the geometry part of LQG the configuration space is 
compact. Secondly, in all the examples considered it so happened that 
$f,T$ did not Poisson commute whenever the spectrum changed. However, 
the idea of \cite{geomop} was to use for $T$ scalar matter while $f$ is 
a geometry function, hence $f,T$ do Poisson commute there.

In \cite{Gambini} we find an example based on $sl(2,R)$ where a 
continuous 
kinematical spectrum switches to a discrete gauge invariant one. 
However, there the configuration space is still non compact and 
furthermore the switch happens in the wrong direction. Thus we now 
explicitly display an example where both of the above 
mentioned properties that are missing in our models are satisfied and 
still the switch occurs.\\
\\    
Consider the phase space $T^\ast(S^1\times R)$ of a particle moving on a 
cylinder. Denote by $A$ the angle configuration variable 
(``connection'') and by $E$ its conjugate momentum (``flux'').
Recall that all geometrical operators of LQG are compound operators 
built from fluxes and they inherit their spectral discreteness from that 
of the flux operators. Denote by $\Phi$ the axis configuration variable
(``scalar field'') and by $\Pi$ its conjugate momentum.
Of course, $A$ is not a globally defined function and we must switch to 
the ``holonomy'' $h=\exp(i A)$. 

We impose as constraint that the particle must spiral around the 
cylinder with period $2\pi$, that is,
\be
\label{B1}
C=A-(\Phi-2\pi [\frac{\Phi}{2\pi}])
\ee
where $[.]$ denotes the Gauss bracket. In the form (\ref{B1}) the 
constraint is not differentiable, hence we pass to the equivalent 
version
\be
\label{B2}
C=\exp(i[A-\Phi])-1=0 \q .
\ee
In this form the constraint is not real valued but that will not pose 
any problems (alternatively work with 
$C=\cos(A-\Phi)-1$).

As clock we choose $T=\Pi$ and as partial observable $f=E$. A 
straightforward calculation reveals that the corresponding complete 
observable is given by
\be \label{B3}
F=E+\Pi-\tau
\ee
which is essentially the total momentum of the particle. 

The kinemtical Hilbert space is given by ${\cal 
H}_{kin}=L_2(S^1,dA/2\pi)\otimes L_2(R,d\Phi)$. Operators are 
represented as $\hat{\Pi}=i d/d\Phi,\;\hat{E}=id/dA=-h \;d/dh$ while 
$\Phi,\;A$ (or rather $h$) become multiplication operators. Clearly, 
$\hat{\Pi},\hat{E},\;\hat{\Phi}$ are self adjoint while $\hat{h}$ is 
unitary. The spectrum of $\hat{E}$ is discrete and takes integer values
with eigenfunctions given by $T_n=h^n$ (``spin network functions'').

To compute the physical Hilbert space we define the rigging map\footnote{The rigging map maps kinematical wave functions to elements of the physical Hilbert space and is used to define the physical inner product. For more details see for instance the last reference of \cite{lqgreviews}.}
\be \label{B4}
\eta(\psi):=\delta(\hat{C})\psi \q .
\ee
The $\delta$--distribution is given by
\be \label{B5}
\delta(\hat{C})=\delta_R(A-(\Phi-2\pi [\frac{\Phi}{2\pi}]))=
\sum_n \; \delta_R(\Phi,A+2\pi n)=\sum_n 
\exp(in[A-\Phi])=\delta_{S^1}(A-\Phi)
\ee
where in the last step we have applied a Poisson resummation.
The physical inner product on the solutions $\eta(\psi)$ is given by
\be \label{B6}
<\eta(\psi),\eta(\psi')>_{phys}=<\psi,\delta(\hat{C})\psi'>_{kin}=
<\tilde{\psi},\tilde{\psi'}>_{L_2(R,d\Phi)}
\ee
where $\tilde{\psi}(\Phi)=\psi(A=\Phi,\Phi)$ which is well defined 
because $\psi$ is periodic in $A$.

One can also get this by making the Fourier expansion
\be \label{B7}
\psi(A,\Phi)=\sum_n T_n(A)\; \psi_n(\Phi)
\ee
and solve $\exp(i[\hat{A}-\hat{\Phi}])\psi=\psi$ for the coefficients
$\psi_n$.

It is 
physically completely obvious that $\hat{F}$ has an absolutely 
continuous spectrum. To see this explicitly, suppose that 
$\eta(\psi)$ is an eigenvector of $\hat{F}$, that is,
\be \label{B8}
\hat{F}\eta(\psi)=\eta(\hat{F}\psi)=\lambda \eta(\psi) \q .
\ee
Since we may assume without loss of generality that $\psi$ depends 
trivially on $A$, this is equivalent to the equation
\be \label{B9}
\hat{\Pi}\psi=[\lambda+\tau]\psi \;\;\Rightarrow \;\; \psi(\Phi)=c\; 
\exp(-i[\lambda+\tau]\Phi)
\ee
for some constant $c$. However, in order to be normalisable in the 
physical inner product it follows $c=0$. Thus $\hat{F}$ has no pure 
point spectrum and since the physical Hilbert space is equivalent to 
$L_2(R,dx)$ on which the observable $\hat{F}$ is essentially represented 
as $id/dx$ it follows that $\hat{F}$ has absolutely continuous spectrum
as claimed. This example illustrates nicely the heuristic argument given in section \ref{section2}: The flow associated to the partial observable $f=E$ is compact on the kinematical phase space, since it integrates to circles. Under the projection to the (gauge fixed and) constraint hypersurface the flow is mapped to non--compact spirals, leading to a continuous spectrum of the associated Dirac observable.

\end{appendix}

\section*{Acknowledgments}

Research at Perimeter Institute for Theoretical Physics is supported in
part by the Government of Canada through NSERC and by the Province of
Ontario through MRI.

\end{document}